\pdfoutput=0
\documentclass[conference]{IEEEtran}
\usepackage{amsmath,amsfonts,amssymb,amsthm,epsfig,epstopdf,url,array}
\usepackage{graphicx}
\usepackage{color}
\usepackage{textcomp,cite}

\begin{document}
\title{Outage Probability Analysis of HARQ-Aided Terahertz Communications}
\author{\IEEEauthorblockN{Ziyang~Song\IEEEauthorrefmark{1}, Zheng~Shi\IEEEauthorrefmark{1}, Qingping Dou\IEEEauthorrefmark{1}, Guanghua~Yang\IEEEauthorrefmark{1}, Yunfei~Li\IEEEauthorrefmark{2}, and 
Shaodan~Ma\IEEEauthorrefmark{2}\\
\IEEEauthorrefmark{1}The School of Intelligent Systems Science and Engineering, Jinan University, China\\
\IEEEauthorrefmark{2}The State Key Laboratory of Internet of Things for Smart City, University of Macau, China}
}
\maketitle
\begin{abstract}
Although terahertz (THz) communications can provide mobile broadband services, it usually has a large path loss and is vulnerable to antenna misalignment. This significantly degrades the reception reliability. To address this issue, the hybrid automatic repeat request (HARQ) is proposed to further enhance the reliability of THz communications. This paper provides an in-depth investigation on the outage performance of two different types of HARQ-aided THz communications, including Type-I HARQ and HARQ with chase combining (HARQ-CC). Moreover, the effects of both fading and stochastic antenna misalignment are considered in this paper. The exact outage probabilities of HARQ-aided THz communications are derived in closed-form, with which the asymptotic outage analysis is enabled to explore helpful insights. In particular, it is revealed that full time diversity can be achieved by using HARQ assisted schemes. Besides, the HARQ-CC-aided scheme performs better than the Type-I HARQ-aided one due to its high diversity combining gain. The analytical results are eventually validated via Monte-Carlo simulations.
\end{abstract}
\begin{IEEEkeywords}
Diversity order, hybrid automatic repeat request (HARQ), outage probability, terahertz (THz) communciations.
\end{IEEEkeywords}
\IEEEpeerreviewmaketitle
\section{Introduction}\label{sec:int}
The past decade has witnessed the explosive growth of wireless devices and the increasing number of bandwidth-consuming internet services, which increase the demand of low-latency high data rate and lead to the spectrum scarcity. To fulfill these challenges, terahertz (THz) communications are envisioned to be a promising wireless technology for beyond fifth generation (B5G) networks, because its wide frequency band allows the data transmission rate up to terabits per second (Tbps) \cite{zhang2021terahertz,tataria20216g,akyildiz20206g}.
Unfortunately, the conventional low-frequency channel model is inapplicable to the THz communications due to its unique propagation characteristics, e.g., atmospheric influence, antenna misalignment, and rain attenuation \cite{liu2021thz}. It has been reported in \cite{boulogeorgos2019analytical,papasotiriou2020performance,ye2021modeling} that these unfavorable factors significantly degrades the reliability performance of THz communications. 
Thus, it is of urgent necessity to address how to enhance the reliability of THz communications.

There are a few works that have been proposed for the reliable THz communications \cite{xia2019expedited,li2018secure,boulogeorgos2019error,lopacinski2018implementation}. In particular,
In \cite{xia2019expedited}, highly directional antennas were proposed to establish reliable THz links. The strategy of integrating THz communications with mobile heterogenous networks was developed to ensure the reliability in \cite{li2018secure}, and the performance of the bit error rate (BER) was then studied. In \cite{boulogeorgos2019error}, a mixed THz-radio frequency wireless architecture was devised to provide a promising reliable backhaul-fronthaul solution. 
Moreover, by noticing that hybrid automatic repeat request (HARQ) is a reliable transmission technique, HARQ assisted THz communications were proposed to investigate the relationship between the BER and the energy efficiency in \cite{lopacinski2018implementation}, in which the discussions are conducted based on system-level simulations. To the best of our knowledge, it lacks of the theoretical analysis for the performance investigation on HARQ-aided THz communications in the literature. To substantially exploit the time diversity gain from HARQ, it is very imperative to study the performance of HARQ-aided THz communications from the information-theoretical perspective. 

Motivated by this, we thoroughly study the outage performance of HARQ-aided THz communications in this paper. The closed-form expressions are derived for the outage probabilities of two types of HARQ-aided THz communications, including Type-I HARQ and HARQ with chase combining (HARQ-CC). With the analytical results, the asymptotic outage analysis is performed to uncover meaningful insights. For example, both Type-I HARQ and HARQ-CC-aided THz schemes can achieve full time diversity. Moreover, the numerical results show that the HARQ-CC-aided THz scheme outperforms the Type-I HARQ-aided one due to its high diversity gain.

The rest of this paper is organized as follows. Section \ref{sec:sys_mod} presents the system model for HARQ-aided THz communications, based on which Section \ref{sec:opa} derives the closed-form expressions for exact and asymptotic outage probabilities for two different types of HARQ-aided THz communications. Section \ref{sec:NR} then shows simulation and numerical results. Finally, Section \ref{sec:con} concludes this paper.
\section{System Model}\label{sec:sys_mod}
In this paper, we consider a point-to-point HARQ-aided THz communication system. To enhance the reception reliability, HARQ technique is utilized to assist THz communications. The received signal at the $k$-th HARQ round can be expressed as

\begin{equation}\label{eqn:channel_model}
{{\bf y}_k} = \sqrt{P}{h_k}{\bf s}_k + {{\bf w}_k},
\end{equation}
where ${P}$ and $h_k$ represents the transmit power and the equivalent THz channel coefficient in the $k$-th HARQ round, respectively, ${\bf s}_k$ is the transmitted symbols with unity power, and ${{\bf w}_k}$ is the complex additive white Gaussian noise (AWGN) with mean zero and variance $N_0$. According to \cite{boulogeorgos2019analytical}, the channel coefficient $h_k$ is modeled according to ${h_k} = {h_l}{h_{pf,k}}$, where $h_l$ is the deterministic THz path gain and keeps constant during all HARQ rounds, and ${h_{pf,k}}$ captures the joint effect of antenna misalignment and multipath fading. From \cite{boulogeorgos2019analytical}, ${h_l}$ is given by
\begin{equation}\label{eqn:path gain}
{h_l} = \frac{{c\sqrt {{G_{t,1}}{G_{r,1}}} }}{{4\pi {f_1}{d_1}}}\exp \left( { - \frac{1}{2}\kappa ({f_1},T,\psi ,p){d_1}} \right),
\end{equation}
where $c$, $f_1$, and $d_1$ stand for the speed of light, the carrier frequency, and the transmission distance, respectively, ${{G_{t,1}}}$ and ${{G_{r,1}}}$ represent the transmit and receive antenna gains, respectively, $\kappa ({f_1},T,\psi ,p)$ characterizes the molecular absorption coefficient, which is determined by the temperature $T$, the relative humidity $\psi$ and the atmospheric pressure $p$. The explicit expression of $\kappa ({f_1},T,\psi ,p)$ is given by \cite[eq. (4)]{boulogeorgos2019analytical}, which is omitted here to conserve space.

Moreover, as proved in \cite{boulogeorgos2019analytical}, the probability density function (PDF) and cumulative distribution function (CDF) of ${h_{pf,k}}$ are respectively expressed as
\begin{equation}\label{eqn:PDF}
{f_{\left| {{h_{pf,k}}} \right|}}(x) = \frac{{\phi {\mu ^{\frac{\phi }{\alpha }}}{x^{\phi  - 1}}}}{{S_0^\phi \hat h_f^\phi \Gamma (\mu )}}\Gamma \left( {\frac{{\alpha \mu  - \phi }}{\alpha },\frac{{\mu {x^\alpha }}}{{S_0^\alpha \hat h_f^\alpha }}} \right),
\end{equation}
\begin{equation}\label{eqn:CDF}
{F_{\left| {{h_{pf,k}}} \right|}}(x) = 1 - \frac{{\phi {\mu ^{\frac{\phi }{\alpha }}}{x^\phi }}}{{\alpha S_0^\phi \hat h_f^\phi }}\sum\limits_{n = 0}^{\mu  - 1} {\frac{1}{{n!}}} \Gamma \left( {\frac{{\alpha n - \phi }}{\alpha },\frac{{\mu {x^\alpha }}}{{S_0^\alpha \hat h_f^\alpha }}} \right),
\end{equation}
where $ \Gamma \left( a,x \right)$ denotes the upper incomplete Gamma function, $\alpha$ is the distribution parameter, $\mu$ and ${\hat h_f}$ denote the variance and the $\alpha$-root mean value of the fading channel envelope, respectively, ${S_0}$ is the fraction of the maximum collected power and is given by ${S_0} = {\left| {{\rm erf}(\zeta )} \right|^2}$ and $\zeta  = \sqrt \pi  {w_{d1}}/\left( {\sqrt 2 {r_1}} \right)$, $r_1$ and ${w_{d1}}$ denote the radius of the receive antenna effective area and the transmission beam footprint radius at reference distance ${d_1}$, respectively, $\phi  = w_e^2/4\sigma _s^2$, ${w_e}$ and ${\sigma _s}$ are the equivalent beam width radius and the doubled spatial jitter standard deviation, respectively, $w_e^2 = \left( {w_{d1}^2\sqrt \pi \rm erf\left( \zeta  \right)} \right)/\left( {2\zeta \exp \left( { - {\zeta ^2}} \right)} \right)$.

To prevent network congestion in the presence of possible deep fading, the maximum number of HARQ rounds is limited up to $M$. In this paper, we consider two types of HARQ schemes, i.e., Type-I HARQ and HARQ-CC. Their difference lies in the diversity combining technique applied at the receiver. More specifically, for Type-I HARQ-aided THz communications, the receiver attempts to recover the message by solely relying on the current packet. The erroneously received packets will be discarded. From the information theoretical perspective \cite{caire2001throughput}, the accumulated mutual information of Type-I HARQ-aided THz communications after $M$ HARQ rounds is given by
\begin{equation}\label{eqn:harq I}
{I^{Type - I}} = \mathop {\max }\limits_{k = 1, \cdots ,M} {\log _2}\left( {1 + \rho {{\left| {{h_{l}}} \right|}^2}{{\left| {{h_{pf,k}}} \right|}^2}} \right),
\end{equation}
where $\rho$ refers to the transmit signal-to-noise ratio (SNR), i.e. $\rho  = {P}/{N_{0}}$.

Furthermore, for HARQ-CC-aided THz communications, the previously failed packets are stored for the combination with the subsequent packet by using maximum ratio combining (MRC). According to \cite{caire2001throughput}, the accumulated mutual information of HARQ-CC-aided THz communications after $M$ HARQ rounds is given by
\begin{equation}\label{eqn:harq cc}
{I^{CC}} = {\log _2}\left( {1 + \sum\limits_{k = 1}^M {\rho {{\left| {{h_{l}}} \right|}^2}{{\left| {{h_{pf,k}}} \right|}^2}} } \right),
\end{equation}
\section{Outage Probability Analysis}\label{sec:opa}
 To study the transmission reliability of HARQ-aided THz wireless systems, the outage probability is the most important performance metric. To be specific, the outage probability is defined as the probability of the event that the accumulated mutual information is less than the transmission rate $R$ \cite{9234486}. Accordingly, the outage probability for HARQ is given by
\begin{equation}\label{eqn:pout}
{P_{out}} = \Pr \left\{ {I < R} \right\},
\end{equation}
where $I \in \left\{ {{I^{Type - I}},{I^{CC}}} \right\}$. By considering different formulations of the accumulated mutual information for the two types of HARQ schemes, the outage analyses for the Type-I and the HARQ-CC-aided THz communications will be undertaken individually in the following subsections.
\subsection{Type-I HARQ-Aided THz Communications}
In what follows, the exact outage analysis will be first carried out for Type-I HARQ-aided THz communications, with which the asymptotic outage analysis is enabled to gain useful insights.
\subsubsection{Exact analysis}
By substituting (\ref{eqn:harq I}) into (\ref{eqn:pout}), ${P_{out}}$ is then expressed as
\begin{equation}\label{eqn:pout_I2}
P_{out}^{Type - I} = \Pr \left\{ {\mathop {\max }\limits_{k = 1, \cdots ,M} {{\log }_2}\left( {1 + \rho {{\left| {{h_{l}}} \right|}^2}{{\left| {{h_{pf,k}}} \right|}^2}} \right) < R} \right\}.
\end{equation}
We assume fast fading channels where the channel coefficients are independent across all HARQ rounds, ${P_{out}}$ can be expressed in terms of the CDF of ${h_{pf,k}}$ as
\begin{align}\label{eqn:pout_II}
P_{out}^{Type - I} &= \prod\limits_{k = 1}^M {\Pr \left\{ {{{\log }_2}(1 + \rho {{\left| {{h_{l}}} \right|}^2}{{\left| {{h_{pf,k}}} \right|}^2}) < R} \right\}} \notag\\
 &= \prod\limits_{k = 1}^M {{F_{\left| {{h_{pf,k}}} \right|}}\left( {\frac{1}{{\left| {{h_{l}}} \right|}}\sqrt {\frac{{{2^R} - 1}}{\rho }} } \right)}.
\end{align}
By substituting the CDF of $h_{pf,k}$ in (\ref{eqn:CDF}) into (\ref{eqn:pout_II}), a closed-form expression of the outage probability of Type-I HARQ-aided THz communications can be expressed as (\ref{eqn:pout_III}), as shown at the top of the next page.
\begin{figure*}[!t]
\normalsize
\begin{equation}\label{eqn:pout_III}
P_{out}^{Type - I} = \prod\limits_{k = 1}^M {\left( {1 - \frac{{\phi {\mu ^{\frac{\phi }{\alpha }}}{{\left( {\sqrt {\frac{{{2^R} - 1}}{\rho }} } \right)}^\phi }}}{{\alpha S_0^\phi \hat h_f^\phi {{\left| {{h_{l}}} \right|}^\phi }}}\sum\limits_{n = 0}^{\mu  - 1} {\frac{1}{{n!}}} \Gamma \left( {\frac{{\alpha n - \phi }}{\alpha },\frac{{\mu {{\left( {\sqrt {\frac{{{2^R} - 1}}{\rho }} } \right)}^\alpha }}}{{S_0^\alpha \hat h_f^\alpha {{\left| {{h_{l}}} \right|}^\alpha }}}} \right)} \right)}
\end{equation}
\hrulefill
\end{figure*}
\subsubsection{Asymptomatic analysis}\label{sec:asy_I}
It is clear that \eqref{eqn:pout_III} is very complicated to extract more meaningful insights. To avoid this, the asymptotic analysis is performed for the outage probability in the high SNR regime, i.e., $\rho  \to \infty $. As proved in Appendix \ref{eqn:8},
the asymptotic outage probability of Type-I HARQ-aided THz communications is given by
\begin{align}\label{eqn:asy I}
&P_{out }^{Type - I} \simeq \notag\\
&\left\{ {\begin{array}{*{20}{c}}
{{{\left( {\frac{{\Gamma \left( {\frac{{\alpha \mu  - \phi }}{\alpha }} \right){\mu ^{\frac{\phi }{\alpha }}}{{\left( {{2^R} - 1} \right)}^{\frac{\phi }{2}}}}}{{\Gamma (\mu )\hat h_f^\phi S_0^\phi {{\left| {{h_l}} \right|}^\phi }{\rho ^{\frac{\phi }{2}}}}}} \right)}^M}}&{,\mu \alpha  - \phi  > 0}\\
{{{\left( {\frac{{\phi {\mu ^{\mu  - 1}}{{\left( {{2^R} - 1} \right)}^{\frac{{\alpha \mu }}{2}}}}}{{\Gamma (\mu )\left( {\phi  - \alpha \mu } \right)\hat h_f^{\alpha \mu }S_0^{\alpha \mu }{{\left| {{h_l}} \right|}^{\alpha \mu }}{\rho ^{\frac{{\alpha \mu }}{2}}}}}} \right)}^M}}&{,\mu \alpha  - \phi  < 0}
\end{array}} \right.
\end{align}

Clearly from \eqref{eqn:asy I}, the diversity order attained by Type I-HARQ-aided THz communications can be obtained, where the diversity order is used to gauge the number of degrees of freedom in the communication system. More precisely, the diversity order is defined as the ratio of the outage probability to the transmit SNR on a log-log scale as \cite{shi2019achievable}
\begin{equation}\label{eqn:div}
d =  - \mathop {\lim }\limits_{\rho  \to \infty } \frac{{\log  {{P_{out}}}}}{{\log \rho  }}.
\end{equation}
By substituting (\ref{eqn:asy I}) into (\ref{eqn:div}), the diversity order of Type-I HARQ-aided THz communications is derived as
\begin{align}\label{eqn:d_I}
{d^{Type - I}} = \left\{ \begin{array}{l}
\frac{{\phi M}}{2},\alpha \mu  - \phi  > 0\\
\frac{{\alpha \mu M}}{2},\alpha \mu  - \phi  < 0
\end{array} \right.
\end{align}
Since the diversity order ${d^{Type - I}}$ is linearly proportional to the maximum number of transmissions, i.e., $M$, full time diversity can be achieved by employing Type-I HARQ.
\subsection{HARQ-CC-Aided THz Communications}
Similarly, the exact and the asymptotic outage analyses are performed for HARQ-CC aided THz communications.
\subsubsection{Exact analysis}
By plugging (\ref{eqn:harq cc}) into (\ref{eqn:pout}) together with some rearrangements, the outage probability ${P_{out}}$ of HARQ-CC aided THz communications can be rewritten as
\begin{equation}\label{eqn:mgf}
P_{out}^{CC} = \Pr \left\{ {\underbrace {\sum\limits_{k = 1}^M {\rho {{\left| {{h_{l}}} \right|}^2}{{\left| {{h_{pf,k}}} \right|}^2}} }_\gamma  < {2^R} - 1} \right\}.
\end{equation}
It is found that $P_{out}^{CC}$ is equivalent to determining the distribution of a summation of multiple random variables, i.e., $\gamma$. Therefore, the method of moment generating function (MGF) is adopted to derive the distribution of $\gamma$. Particularly, the MGF of the PDF of $\gamma$ can be obtained as
\begin{align}\label{eqn:mgff}
{\mathbb E_\gamma }\left\{ {{e^{t\gamma }}} \right\} =& \prod\limits_{k = 1}^M {\mathbb E\left\{ {{e^{\rho {{\left| {{h_{l}}} \right|}^2}{{\left| {{h_{pf,k}}} \right|}^2}t}}} \right\}} \notag\\
 =& \prod\limits_{k = 1}^M {\frac{{\phi {\mu ^{\frac{\phi }{\alpha }}}}}{{S_0^\phi \hat h_f^\phi \Gamma (\mu )}}\int_{\rm{0}}^\infty  {{x_k}^{\phi  - 1}{e^{\rho {{\left| {{h_{l}}} \right|}^2}{x_k}^2t}}}}\notag\\
 &\times\Gamma \left( {\frac{{\alpha \mu  - \phi }}{\alpha },\frac{{\mu x_k^\alpha }}{{S_0^\alpha \hat h_f^\alpha }}} \right)  d{x_k},
\end{align}
where $\mathbb E\{\cdot\}$ stands for the expectation operator. Then, the CDF of $\gamma $ can be obtained by inverse Laplace transform as
\begin{equation}\label{eqn:lap}
{F_\gamma }(\gamma ) = \frac{1}{{2\pi \rm i}}\int_{\alpha  - \rm i\infty }^{\alpha {\rm{ + \rm i}}\infty } {{e^{\gamma t}}\frac{1}{t}{\mathbb E_\gamma }\left\{ {{e^{ - t\gamma }}} \right\}dt}.
\end{equation}
where $\alpha  > 0$ and $\rm i = \sqrt { - 1}$. By substituting (\ref{eqn:mgff}) into (\ref{eqn:lap}), it follows that
\begin{align}\label{eqn:lapp}
&{F_\gamma }(\gamma ) = \frac{1}{{2\pi \rm i}}\int_{\alpha  - \rm i\infty }^{\alpha {\rm{ + \rm i}}\infty } {{e^{\gamma t}}\frac{1}{t}\prod\limits_{k = 1}^M {\frac{{\phi {\mu ^{\frac{\phi }{\alpha }}}}}{{S_0^\phi \hat h_f^\phi \Gamma (\mu )}}}}\notag\\
&\times\int_{\rm{0}}^\infty  {{x_k}^{\phi  - 1}{e^{ - \rho {{\left| {{h_{l}}} \right|}^2}{x_k}^2t}}\Gamma \left( {\frac{{\alpha \mu  - \phi }}{\alpha },\frac{{\mu x_k^\alpha }}{{S_0^\alpha \hat h_f^\alpha }}} \right)}  d{x_k}dt.
\end{align}
By applying Parseval's Type Property of Mellin transform \cite[Eq. 8.3.22]{debnath2014integral} to the inner integral of \eqref{eqn:lapp}, one has
\begin{align}\label{eqn:Parr}
{F_\gamma }(\gamma ) =& \frac{1}{{2\pi \rm i}}\int_{\alpha  {\rm{ - i}}\infty }^{\alpha {\rm{ + i}}\infty } {{e^{\gamma t}}\frac{1}{t}\prod\limits_{k = 1}^M {\frac{\phi }{{2\hat h_f^{\alpha  - \phi }\Gamma (\mu )}}}} \notag\\
&\times\frac{1}{{2\pi \rm i}}\int_{\alpha  {\rm{ - i}}\infty }^{\alpha {\rm{ + i}}\infty } {\frac{{\Gamma \left( {\frac{s}{2}} \right)\Gamma \left( {\phi  - s} \right)\Gamma \left( {\mu  - \frac{s}{\alpha }} \right)}}{{\Gamma \left( {1 + \phi  - s} \right)}}}\notag\\
&\times{{\left( {{{\left( {\rho {{\left| {{h_{l}}} \right|}^2}t} \right)}^{\frac{1}{2}}}{{\left( {\frac{\mu }{{\hat h_f^\alpha S_{\rm{0}}^\alpha }}} \right)}^{ - \frac{1}{\alpha }}}} \right)}^{ - s}}ds dt,
\end{align}
where \eqref{eqn:Parr} holds by using the following two Mellin transforms \cite[Eq. 8.2.5]{debnath2014integral}, \cite[Eq. 6.455, Eq. 3.326.2]{gradshteyn2014table} 
\begin{align}\label{eqn:for}
& \int_{\rm{0}}^\infty  {{x_k}^{s - 1}{e^{ - \rho {{\left| {{h_{l}}} \right|}^2}x_k^2 t}}d} {x_k}= \frac{1}{2}{\left( {\rho {{\left| {{h_{l}}} \right|}^2}t} \right)^{ - \frac{s}{2}}}\Gamma \left( {\frac{s}{2}} \right),
\end{align}
\begin{multline}\label{eqn:for_2}
 \int_{\rm{0}}^\infty  {{x_k}^{s - 1}\Gamma \left(\frac{{\alpha\mu - \phi }}{\alpha },\frac{{\mu x_k^\alpha }}{{S_{\rm{0}}^\alpha h_f^\alpha }}\right)d} {x_k}\\
 = \frac{1}{{s}}{\left( {\frac{\mu}{{h_f^\alpha S_{\rm{0}}^\alpha }}} \right)^{ - \frac{{s}}{\alpha }}}\Gamma \left(\frac{{\alpha\mu + s - \phi}}{\alpha }\right).
\end{multline}

By recognizing the inner integral as a Mellin-Barnes integral and identifying it with the extended Fox's H function, we derive ${F_\gamma }(\gamma )$ in a compact form, where the definition of the extended Fox's H function is given by (\ref{eqn:HF}) \cite[Eq.T.I.1]{ansari2017new}, as shown at the top of the next page.
\begin{figure*}[!t]
\normalsize
\begin{equation}\label{eqn:HF}
 H_{p,q}^{m,n}\left[ {z|_{{{\left( {{\beta _j},{B_{j,}}{b_j}} \right)}_{1,q}}}^{{{\left( {{\alpha _j},{A_j},{a_j}} \right)}_{1,p}}}} \right] = \frac{1}{{2\pi \rm i}}\oint_C {\frac{{\prod\nolimits_{j = 1}^n {{{\left( {\Gamma \left( {1 - {\alpha _j} + {A_j}s} \right)} \right)}^{{a_j}}}\prod\nolimits_{j = 1}^m {{{\left( {\Gamma \left( {{\beta _j} - {B_j}s} \right)} \right)}^{{b_j}}}} } }}{{\prod\nolimits_{j = n + 1}^p {{{\left( {\Gamma \left( {{\alpha _j} - {A_j}s} \right)} \right)}^{{a_j}}}\prod\nolimits_{j = m + 1}^q {{{\left( {\Gamma \left( {1 - {\beta _j} + {B_j}s} \right)} \right)}^{{b_j}}}} } }}} {z^s}ds,
\end{equation}
\end{figure*}
As a consequence, by substituting $\gamma  = {2^R} - 1$ into ${F_\gamma }(\gamma )$, we arrive at the final expression of the outage probability of HARQ-CC aided THz communications as (\ref{eqn:harq_ccc}), as shown at the top of the next page.
\begin{figure*}[!t]
\normalsize
\begin{equation}\label{eqn:harq_ccc}
P_{out}^{CC} = {F_\gamma }({2^R} - 1) = {\left( {\frac{\phi }{{2\Gamma (\mu)}}} \right)^M}\frac{1}{{2\pi \rm i}}\int_{\alpha  - \rm i\infty }^{\alpha {\rm{ + \rm i}}\infty } {\frac{{{e^{\left( {{2^R} - 1} \right)t}}}}{t}\prod\limits_{k = 1}^M {H_{0,1}^{1,2}\left[ {{{\left( {\rho {{\left| {{h_{l}}} \right|}^2}t} \right)}^{\frac{1}{2}}}{{\left( {\frac{\mu}{{h_f^\alpha S_{\rm{0}}^\alpha }}} \right)}^{ - \frac{1}{\alpha }}}\left| {_{\left( {0,\frac{1}{2}} \right),\left( { - \phi ,1} \right)}^{\left( {1 - \phi ,1} \right),\left( {1 - \mu,\frac{1}{\alpha }} \right)}} \right.} \right]} dt},
\end{equation}
\hrulefill
\end{figure*}

\subsubsection{Asymptomatic analysis}
Similarly to Section \ref{sec:asy_I}, the asymptomatic outage analysis is conducted for (\ref{eqn:harq_ccc}) under high SNR, i.e., $\rho  \to \infty $. As proved in Appendix \ref{eqn:CC}, as the transmit SNR $\rho $ approaches to infinity, the outage probability is asymptotic to
\begin{align}\label{eqn:asy cc}
&P_{out}^{CC} \simeq  \eta P_{out }^{Type - I},
\end{align}
where $\eta$ is given by
\begin{equation}\label{eqn:eta}
\eta  = \left\{ {\begin{array}{*{20}{c}}
{\frac{{{{\left( {\Gamma \left( {\frac{\phi }{2} + 1} \right)} \right)}^M}}}{{\Gamma \left( {\frac{{\phi M}}{2} + 1} \right)}},}&{\mu\alpha  - \phi  > 0}\\
{\frac{{{{\left( {\Gamma \left( {\frac{{\mu \alpha }}{2} + 1} \right)} \right)}^M}}}{{\Gamma \left( {\frac{{\mu \alpha M}}{2} + 1} \right)}},}&{\mu\alpha  - \phi  < 0}
\end{array}} \right.
\end{equation}

By substituting (\ref{eqn:asy cc}) into (\ref{eqn:div}), the diversity order of HARQ-CC-aided THz communications is the same as Type-I HARQ-aided ones, i.e.,
\begin{align}\label{eqn:div_2}
 {d^{CC}} = {d^{Type - I}}.
\end{align}
This indicates that both the two HARQ-aided THz communication schemes achieve the full time diversity. Moreover, it is not hard to prove that $\eta$ is less than 1 for $M>1$. From \eqref{eqn:asy cc}, this result justifies that the outage performance of HARQ-CC aided THz communications outperforms that of Type-I HARQ-aided THz communications due to its high diversity combining gain.
\section{Numerical Results}\label{sec:NR}
In this section, the simulation results are presented to verify the analytical results. Unless otherwise stated, the system parameters are set as follows, $\alpha  = 2$, $\mu  = 1$, $\psi  = 50\% $, $T = 296{\rm{^\circ }}$~K, $p = 101325$~Pa, $M = 3$, ${\sigma _{\rm{s}}} = 1$ and the carrier frequency is $f_1=275$~THz. Moreover, both the transmit and the receive antenna gains equal to $55$~dBi, the transmission distance is assumed to be $20$~m. By considering the effect of the sign of $\alpha \mu  - \phi$ on the asymptotic analysis, we assume that the equivalent beamwidth ${w_e}$ is set as $1$ for $\alpha \mu  - \phi  > 0$, while ${w_e}$ is set as $3$ for $\alpha \mu  - \phi  < 0$.

\begin{figure}
        \centering
        \includegraphics[width=3.5in]{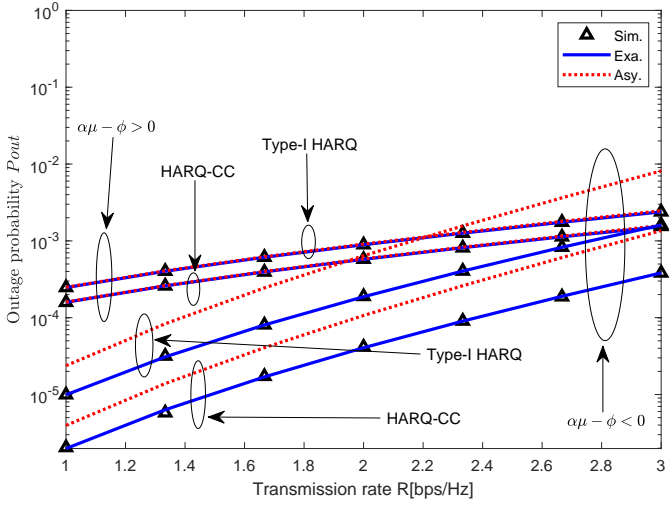}
        \caption{The outage probability $P_{out}$ versus the transmission rate $R$ with $\rho = 30$~dB.}\label{fig:R2}
\end{figure}

Fig. \ref{fig:R2} investigates the relationship between the outage probability ${P_{out}}$ and the transmission rate $R$, where the labels 'Sim.', 'Exa.', and 'Asy.' represent the simulated, the exact and the asymptotic outage probabilities, respectively. It can be seen that the exact results perfectly coincide with the simulated results. Moreover, it is as expected that the outage probability increases with transmission rate. Furthermore, it is easily observed that HARQ-CC-aided THz communications is superior to Type-I HARQ-aided THz communications due to the high diversity combining gain no matter if $\alpha \mu  - \phi  > 0$ or not. This observation clearly confirms the validity of our asymptotic results. 

\begin{figure}
        \centering
        \includegraphics[width=3.5in]{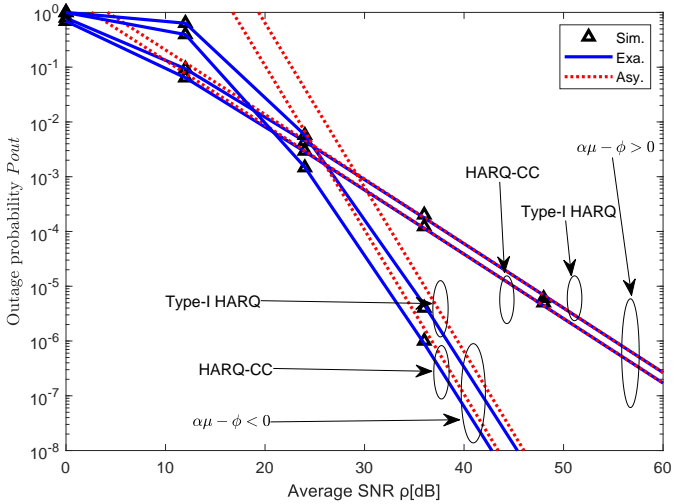}
        \caption{The outage probability $P_{out}$ versus the average SNR $\rho $ with $R = 2$~bps/Hz.}\label{fig:R3}
\end{figure}

Fig. \ref{fig:R3} depicts the relationship between the outage probability ${P_{out}}$ and the average transmit SNR for the two HARQ-aided THz communication schemes. Similarly, the exact and the simulated results match well with each other, and they converge to the asymptotic ones in the high SNR regime no matter whether $\alpha \mu  - \phi  > 0$ or not. Besides, it is found that the decreasing slopes of both schemes are the same, that is, the outage curves of the two schemes are parallel to each other. This observation is consistent with the analysis of the diversity order. 
\section{Conclusions}\label{sec:con}
In this paper, we investigated the outage performance of two different types of HARQ-aided THz communications, including Type-I HARQ and HARQ-CC. The main contribution of this paper was devoted to quantify the effect of the fading and antenna misalignment on the outage performance through the exact and the asymptotic analyses. The results filled the gap for the performance analysis of HARQ-aided THz communications from information-theoretical perspective. It revealed that the Type-I HARQ-aided THz communication scheme are inferior to the HARQ-CC-aided one due to its low diversity combining gain, but the analysis of the diversity order indicated that both Type-I HARQ and HARQ-CC-aided schemes can achieve full time diversity. 
The analytical results were finally validated through Monte Carlo simulations. 

\appendices
\section{Proof of \eqref{eqn:asy I}}\label{eqn:8}
With the PDF of $|{h_{pf,k}}|$ given by \eqref{eqn:PDF}, (\ref{eqn:pout_II}) can be rewritten as
\begin{align}\label{eqn:20}
&P_{out}^{Type - I} = \prod\limits_{k = 1}^M {\int\limits_0^{\frac{1}{{\left| {{h_{l}}} \right|}}\sqrt {\frac{{{2^R} - 1}}{\rho }} } {{f_{\left| {{h_{pf,k}}} \right|}}(x)dx} }\notag\\
&= \prod\limits_{k = 1}^M {\int\limits_0^{\frac{1}{{\left| {{h_{l}}} \right|}}\sqrt {\frac{{{2^R} - 1}}{\rho }} } {\frac{{\phi {\mu ^{\frac{\phi }{\alpha }}}{x^{\phi  - 1}}}}{{S_0^\phi \hat h_f^\phi \Gamma (\mu )}}\Gamma \left( {\frac{{\alpha \mu  - \phi }}{\alpha },\frac{{\mu {x^\alpha }}}{{S_0^\alpha \hat h_f^\alpha }}} \right)dx} }.
\end{align}
By noticing that the upper incomplete Gamma function in \eqref{eqn:20} has different asymptotic behaviors for $\alpha \mu  - \phi  > 0$ and $\alpha \mu  - \phi  < 0$, the asymptotic expressions of the outage probability in these two cases are derived one by one.
\subsection{$\alpha \mu  - \phi  > 0$}
By using the first mean value theorem \cite[Eq. 1.4.29]{olver2010nist}, we have
\begin{align}\label{eqn:Mean}
&P_{out}^{Type - I} = \prod\limits_{k = 1}^M {\Gamma \left( {\frac{{\alpha \mu  - \phi }}{\alpha },\frac{{\mu {\varsigma ^\alpha }}}{{S_0^\alpha \hat h_f^\alpha }}} \right)\int\limits_0^{\frac{1}{{\left| {{h_{l}}} \right|}}\sqrt {\frac{{{2^R} - 1}}{\rho }} } {\frac{{\phi {\mu ^{\frac{\phi }{\alpha }}}{x^{\phi  - 1}}}}{{S_0^\phi \hat h_f^\phi \Gamma (\mu )}}dx} } \notag\\
 =& \prod\limits_{k = 1}^M {\Gamma \left( {\frac{{\alpha \mu  - \phi }}{\alpha },\frac{{\mu {\varsigma ^\alpha }}}{{S_0^\alpha \hat h_f^\alpha }}} \right)\frac{{{\mu ^{\frac{\phi }{\alpha }}}}}{{S_0^\phi \hat h_f^\phi \Gamma (\mu )}}}{{\left( {\frac{1}{{\left| {{h_{l}}} \right|}}\sqrt {\frac{{{2^R} - 1}}{\rho }} } \right)}^\phi },
\end{align}
where $\varsigma  \in \left[ {0,\sqrt {{({{2^R} - 1})}/{\rho }}/{{\left| {{h_{l}}} \right|}} } \right]$. In addition, it is found in the high SNR regime that $\varsigma$ tends to $0$ for $\rho \to \infty$. Thus, if $\alpha \mu  - \phi  > 0$, $\Gamma \left( {\left( {\alpha \mu  - \phi } \right){\alpha ^{ - 1}},\mu {\varsigma ^\alpha }\hat h_f^{ - \alpha }S_0^{ - \alpha }} \right)$ approaches to 
$\Gamma \left( {\left( {\alpha \mu  - \phi } \right){\alpha ^{ - 1}}} \right)$. 
As a consequence, the asymptotic expression of the outage probability in the case of $\alpha \mu  - \phi  > 0$ can be obtained as \eqref{eqn:asy I}.
\subsection{$\alpha \mu  - \phi  < 0$}
By using the asymptotic expression of the upper incomplete gamma function $ \Gamma \left( a,x \right)\simeq -x^a/a$ for $x\to 0$ if $a<0$ \cite[Eq. 8.7.3]{olver2010nist}, the asymptotic outage probability in the high SNR regime can be obtained as
\begin{align}\label{eqn:23}
&P_{out }^{Type - I} \notag\\
&= \prod\limits_{k = 1}^M {\int\limits_0^{\frac{1}{{\left| {{h_{l}}} \right|}}\sqrt {\frac{{{2^R} - 1}}{\rho }} } {\frac{{\phi {\mu ^{\frac{\phi }{\alpha }}}{x^{\phi  - 1}}}}{{S_0^\phi \hat h_f^\phi \Gamma (\mu )}}\frac{{\alpha {{\left( {\frac{\mu }{{\hat h_f^\alpha S_0^\alpha }}} \right)}^{\frac{{\alpha \mu  - \phi }}{\alpha }}}}}{{\phi  - \alpha \mu }}{x^{\alpha \mu  - \phi }}dx} }\notag\\
 &= {{{\left( {\frac{{\phi {\mu ^{\mu  - 1}}{{\left( {{2^R} - 1} \right)}^{\frac{{\alpha \mu }}{2}}}}}{{\left( {\phi  - \alpha \mu } \right)S_0^{\alpha \mu }\widehat h_f^{\alpha \mu }{\rho ^{\frac{{\alpha \mu }}{2}}}\Gamma (\mu ){{\left| {{h_l}} \right|}^{\alpha \mu }}}}} \right)}^M}}.
\end{align}
Finally, the asymptotic expressions of the outage probability for Type I-HARQ-aided THz communications can be derived as \eqref{eqn:asy I}.
\section{Proof of \eqref{eqn:asy cc}}\label{eqn:CC}
To obtain the asymptotic expression of \eqref{eqn:harq_ccc} under high SNR, it is obligatory to derive the asymptotic expression of the extended Fox's H function for $\rho\to \infty$. By using the residue theorem, the extended Fox's H function in \eqref{eqn:harq_ccc} is asymptotic to \eqref{eqn:res}, as shown at the top of the next page, where ${\mathop {{\rm{Res}}}\limits_{s = a} }\{f(s)\}$ denotes the residue of $f(s)$ at $s=a$ and the second step holds by using dominant term approximation under the condition of $\rho\to \infty$.
\begin{figure*}[!t]
\begin{align}\label{eqn:res}
&H_{0,1}^{1,2}\left[ {\left. {{{\left( {\rho {{\left| {{h_l}} \right|}^2}t} \right)}^{\frac{1}{2}}}{{\left( {\frac{\mu }{{h_f^\alpha S_{\rm{0}}^\alpha }}} \right)}^{ - \frac{1}{\alpha }}}} \right|\begin{array}{*{20}{c}}
{\left( {1 - \phi ,1} \right),\left( {1 - \mu ,\frac{1}{\alpha }} \right)}\\
{\left( {0,\frac{1}{2}} \right),\left( { - \phi ,1} \right)}
\end{array}} \right] \notag\\
&=  - \sum\nolimits_{\scriptstyle\left\{ {a:a = \phi  + n,n \in [0,\infty ]} \right\}\hfill\atop
\scriptstyle\bigcup {\left\{ {a:a = \mu \alpha  + n\alpha ,n \in [0,\infty ]} \right\}} \hfill} {\mathop {{\rm{Res}}}\limits_{s = a} \left\{ {\frac{{\Gamma \left( {\frac{s}{2}} \right)\Gamma \left( {\phi  - s} \right)\Gamma \left( {\mu  - \frac{s}{\alpha }} \right)}}{{\Gamma \left( {1 + \phi  - s} \right)}}{{\left( {{{\left( {\rho {{\left| {{h_l}} \right|}^2}t} \right)}^{\frac{1}{2}}}{{\left( {\frac{\mu }{{h_f^\alpha S_{\rm{0}}^\alpha }}} \right)}^{ - \frac{1}{\alpha }}}} \right)}^{ - s}}} \right\}} \notag\\
 &\simeq  - \mathop {\lim }\limits_{s \to \phi } (s - \phi )\frac{{\Gamma \left( {\frac{s}{2}} \right)\Gamma \left( {\phi  - s} \right)\Gamma \left( {\mu  - \frac{s}{\alpha }} \right)}}{{\Gamma \left( {1 + \phi  - s} \right)}}{\left( {{{\left( {\rho {{\left| {{h_l}} \right|}^2}t} \right)}^{\frac{1}{2}}}{{\left( {\frac{\mu }{{h_f^\alpha S_{\rm{0}}^\alpha }}} \right)}^{ - \frac{1}{\alpha }}}} \right)^{ - s}}\notag\\
 &- \mathop {\lim }\limits_{s \to \mu \alpha } (s - \mu \alpha )\frac{{\Gamma \left( {\frac{s}{2}} \right)\Gamma \left( {\phi  - s} \right)\Gamma \left( {\mu  - \frac{s}{\alpha }} \right)}}{{\Gamma \left( {1 + \phi  - s} \right)}}{\left( {{{\left( {\rho {{\left| {{h_l}} \right|}^2}t} \right)}^{\frac{1}{2}}}{{\left( {\frac{\mu }{{h_f^\alpha S_{\rm{0}}^\alpha }}} \right)}^{ - \frac{1}{\alpha }}}} \right)^{ - s}}\notag\\
 &={\left( {\rho t} \right)^{ - \frac{\phi }{2}}}\underbrace {\Gamma \left( {\frac{\phi }{2}} \right)\Gamma \left( {\mu - \frac{\phi }{\alpha }} \right){{\left( {{{\left( {{{\left| {{h_{l}}} \right|}^2}} \right)}^{\frac{1}{2}}}{{\left( {\frac{\mu}{{h_f^\alpha S_{\rm{0}}^\alpha }}} \right)}^{ - \frac{1}{\alpha }}}} \right)}^{ - \phi }}}_{{B_k}} + {\left( {\rho t} \right)^{ - \frac{{\mu\alpha }}{2}}}\underbrace {\frac{\alpha }{{\phi  - \mu\alpha }}\Gamma \left( {\frac{{\mu\alpha }}{2}} \right){{\left( {{{\left( {{{\left| {{h_{l}}} \right|}^2}} \right)}^{\frac{1}{2}}}{{\left( {\frac{\mu}{{h_f^\alpha S_{\rm{0}}^\alpha }}} \right)}^{ - \frac{1}{\alpha }}}} \right)}^{ - \mu\alpha }}}_{{C_k}},
\end{align}
\hrulefill
\end{figure*}
By substituting \eqref{eqn:res} into \eqref{eqn:harq_ccc} and by change of variable as by $y = \rho t$, it leads to
\begin{multline}\label{eqn:pt}
P_{out}^{CC} 
\simeq {\left( {\frac{\phi }{{2\Gamma (\mu)}}} \right)^M}\frac{1}{{2\pi \rm i}}\int_{\alpha  {\rm{ - i}}\infty }^{\alpha {\rm{ + i}}\infty } {\frac{{{e^{\left( {{2^R} - 1} \right)\frac{y}{\rho }}}}}{y}}\\
\times\prod\limits_{k = 1}^M {\left( {{B_k}{y^{ - \frac{\phi }{2}}} + {C_k}{y^{ - \frac{{\mu\alpha }}{2}}}} \right)} dy.
\end{multline}
\eqref{eqn:pt} can be further expanded as
\begin{align}\label{eqn:cc_1}
P_{out}^{CC} &\simeq {\left( {\frac{\phi }{{2\Gamma (\mu)}}} \right)^M}\frac{1}{{2\pi \rm i}}\int_{\alpha  {\rm{ - i}}\infty }^{\alpha {\rm{ + i}}\infty } {\frac{{{e^{\left( {{2^R} - 1} \right)\frac{y}{\rho }}}}}{y}}\notag\\
&\times\sum\limits_{{\bf{a}} \in {\bf{\Omega }}} {\prod\limits_{k = 1}^M {{{\left( {{B_k}{y^{ - \frac{\phi }{2}}}} \right)}^{{a_k}}}{{\left( {{C_k}{y^{ - \frac{{\mu\alpha }}{2}}}} \right)}^{1 - {a_k}}}} } dy \notag\\
 &= {\left( {\frac{\phi }{{2\Gamma (\mu)}}} \right)^M}\sum\limits_{{\bf{a}} \in {\bf{\Omega }}} {\prod\limits_{k = 1}^M {{B_k}^{{a_k}}{C_k}^{1 - {a_k}}} } \notag\\
 &\times\frac{1}{{2\pi \rm i}}\int_{\alpha  {\rm{ - i}}\infty }^{\alpha {\rm{ + i}}\infty } {{e^{\left( {{2^R} - 1} \right)\frac{y}{\rho }}}{y^{ - \sum\limits_{k = 1}^M {\left( {\frac{{\mu\alpha }}{2}\left( {1 - {a_k}} \right) + \frac{\phi }{2}{a_k}} \right)}  - 1}}dy},
\end{align}
where ${\bf{\Omega }} = \left\{ { {\left( {{a_1}, \cdots ,{a_M}} \right)} : {a_k} = \left\{ {0,1} \right\},1 \le k \le M} \right\}$. By means of inverse Laplace transform \cite[Eq. 3.382.7]{gradshteyn2014table}, \eqref{eqn:cc_1} can be simplified as
\begin{align}\label{eqn:cc_2}
P_{out}^{CC} 
 \simeq& {\left( {\frac{\phi }{{2\Gamma (\mu)}}} \right)^M}\sum\limits_{{\bf{a}} \in {\bf{\Omega }}} {\frac{{\prod\limits_{k = 1}^M {{B_k}^{{a_k}}{C_k}^{1 - {a_k}}} }}{{\Gamma \left( {{M\frac{{\mu\alpha }}{2} + {\frac{{\phi  - \mu\alpha }}{2}} \sum\limits_{k = 1}^M {{a_k}} }  + 1} \right)}}}\notag\\
 &\times{{\left( {\frac{{{2^R} - 1}}{\rho }} \right)}^{M\frac{{\mu\alpha }}{2} + {\frac{{\phi  - \mu\alpha }}{2}} \sum\limits_{k = 1}^M {{a_k}} }}.
\end{align}
In the high SNR regime, i.e., $\rho\to \infty$, to get the dominant terms $\rho^{-{\left(M\frac{{\mu\alpha }}{2} + {\frac{{\phi  - \mu\alpha }}{2}} \sum\nolimits_{k = 1}^M {{a_k}} \right)}}$ amounts to finding the minimum value of the corresponding exponent ${M\frac{{\mu\alpha }}{2} +  {\frac{{\phi  - \mu\alpha }}{2}} \sum\nolimits_{k = 1}^M {{a_k}} }$. Clearly, the minimum of the exponent depends on the sign of ${\phi  - \mu\alpha }$. More specifically, the minimum of the exponent is attained with $a_1=\cdots=a_M=0$ if ${\phi  - \mu\alpha }>0$ and $a_1=\cdots=a_M=1$ otherwise. With this result, \eqref{eqn:cc_2} can be finally asymptotically expressed as
\begin{align}\label{eqn:cc_int}
&P_{out}^{CC} \simeq \notag\\
& \left\{ {\begin{array}{*{20}{c}}
{\frac{{{{\left( {\Gamma \left( {\frac{\phi }{2} + 1} \right)} \right)}^M}}}{{\Gamma \left( {\frac{{\phi M}}{2} + 1} \right)}}{{\left( {\frac{{\Gamma \left( {\frac{{\alpha \mu  - \phi }}{\alpha }} \right){\mu ^{\frac{\phi }{\alpha }}}{{\left( {{2^R} - 1} \right)}^{\frac{\phi }{2}}}}}{{\Gamma (\mu )\hat h_f^\phi S_0^\phi {{\left| {{h_l}} \right|}^\phi }{\rho ^{\frac{\phi }{2}}}}}} \right)}^M},}&{\mu\alpha  - \phi  > 0}\\
{\frac{{{{\left( {\Gamma \left( {\frac{{\mu\alpha }}{2} + 1} \right)} \right)}^M}}}{{\Gamma \left( {\frac{{\mu\alpha M}}{2} + 1} \right)}}{{\left( {\frac{{\phi {\mu ^{\mu  - 1}}{{\left( {{2^R} - 1} \right)}^{\frac{{\alpha \mu }}{2}}}}}{{\Gamma (\mu )\left( {\phi  - \alpha \mu } \right)\hat h_f^{\alpha \mu }S_0^{\alpha \mu }{{\left| {{h_l}} \right|}^{\alpha \mu }}{\rho ^{\frac{{\alpha \mu }}{2}}}}}} \right)}^M},}&{\mu\alpha  - \phi  < 0}
\end{array}} \right.
\end{align}
By comparing \eqref{eqn:cc_int} with \eqref{eqn:asy I}, the asymptotic outage probability of HARQ-CC-aided THz communications can be rewritten as \eqref{eqn:asy cc} in the end.
\bibliographystyle{ieeetran}
\bibliography{THz}
\end{document}